\documentclass [a4paper,fleqn, 12pt]{article}
\usepackage{graphicx}
\usepackage[small]{subfigure,epsfig}

\usepackage {amsmath} \usepackage{amssymb} \usepackage{cite}

\numberwithin{equation}{section} \numberwithin{table}{section}
\newcommand{\cn}

\begin{document}


\title
{Point vortices and polynomials of the Sawada -- Kotera and Kaup -- Kupershmidt equations}

\author
{Maria V Demina,  \and   Nikolay A. Kudryashov}

\date{Department of Applied Mathematics, National Research Nuclear University
MEPHI, 31 Kashirskoe Shosse,
115409 Moscow, Russian Federation}




\maketitle

\begin{abstract}

Rational solutions and special polynomials associated with the generalized $K_2$ hierarchy are studied. This hierarchy is related to the Sawada -- Kotera and Kaup -- Kupershmidt equations and some other integrable partial differential equations including the Fordy -- Gibbons equation. Differential -- difference relations and differential equations satisfied by the polynomials are derived. The relationship between these special polynomials
and stationary configurations of point vortices with circulations $\Gamma$ and $-2 \Gamma$ is established. Properties of the polynomials are studied. Differential -- difference relations enabling one to construct these polynomials explicitly are derived. Algebraic relations satisfied by the roots of the polynomials are found.

\end{abstract}






\section{Introduction}

Vortical flows belong to one of the most important issues of fluid dynamics \cite{Tomson01, Tomson02, Kadtke01, Campbell01, Campbell02, Campbell03, Borisov01}. The motion of point vortices including the problem of finding
equilibrium vortex configurations has been intensively studied in recent years \cite{Aref01, Aref02, Oneil01, Oneil02, Aref03, Aref04, Clarkson01, Oneil03}. It is a well--known fact that positions of $N$ identical point vortices
in equilibrium on a line are described by the roots of the $N^{\text{th}}$ Hermite polynomial \cite{Aref01}. According to another classical result $N$ identical point vortices
in equilibrium on a circle form a regular $N$--gon \cite{Aref01,Aref04, Oneil03}.

Not long ago a striking relationship between the theory of point vortices and rational solutions of some integrable partial differential equations
was established. Stationary equilibria of point vortices with circulations $\Gamma$ and $-\Gamma$ are described by the roots of the Adler -- Moser polynomials \cite{Kadtke01, Clarkson01, Adler01}.
Originally these polynomials were introduced to construct rational solutions of the Korteweg --- de Vries equation. The Adler -- Moser polynomials at certain values of the parameters are known as
the Yablonskii --- Vorob'ev polynomials, which are used to represent rational solutions of the second Painlev\'{e} equation \cite{Clarkson02, Demina15, Kudr08a}. The third, the fourth, the fifth, and the sixth Painlev\'{e} equations also possess rational or algebraic solutions expressible in terms of certain special polynomials \cite{Clarkson02}. For example, rational solutions of the fourth  Painlev\'{e} equation can be expressed via logarithmic derivative of the generalized Hermite and the generalized Okamoto polynomials \cite{Clarkson01, Clarkson02}.

Special polynomials associated with the Painlev\'{e} equations and their higher--order analogues
 have been attracting much attention during recent decades. It was shown that these polynomials possess a certain number of interesting properties. For example, their roots form highly regular structures in the complex plane.

In this article we investigate the connection between equilibria of point vortices and special polynomials associated with
rational solutions of the generalized $K_2$ hierarchy. This hierarchy is related to the Sawada~-- Kotera equation \cite{Sawada},  the Kaup -- Kupershmidt equation \cite{Kupershmidt}, their hierarchies \cite{Caudrey01, Weiss01}, and some other integrable partial differential equations including the Fordy -- Gibbons equation \cite{Fordy01}.

This article is organized as follows. In section \ref{Dynamics} we consider equilibria of point vortices with circulations $\Gamma$ and~$-\mu \Gamma$, $\mu>0$. In section \ref{Gen_K2} we study properties of special polynomials associated with the generalized $K_2$ hierarchy, we derive differential -- difference relations and ordinary differential equations satisfied by the polynomials. In section \ref{Gen_K2_1} we discuss the case of the fourth--order equation in the generalized $K_2$ hierarchy.

\section{Dynamics of point vortices} \label{Dynamics}

One of the basic systems of equations in the theory of point vortices can be written as
\begin{equation}
\label{Motion_of_Vortices}\frac{d z_k^{*}}{d\,t}=\frac{1}{2\pi i}\sum_{j=1}^{M}{}^{'}\frac{\Gamma_j}{z_k-z_j},\quad k=1,\ldots, M.
\end{equation}
These equations describe the motion of $M$ point vortices with circulations (or strengths)
$\Gamma_k$ at positions $z_k$, $k=1$, $\ldots$, $M$. The prime in expression \eqref{Motion_of_Vortices} means that we exclude the case $j=k$ and the symbol~$\,^{*}$ stands for complex conjugation. In this article we study stationary equilibria of vortices, thus we set $d z_k^{*}/ dt =0$.
First of all, let us consider the situation with $l_1$ vortices of circulation $\Gamma$ at positions $a_1$, $\ldots$, $a_{l_1}$ and
$l_2$ vortices of circulation $-\mu \Gamma$, $\mu>0$ at positions $b_1$, $\ldots$, $b_{l_2}$.
For further analysis it is convenient to introduce the polynomials $P(z)$ and $Q(z)$ with roots at the positions of vortices \cite{Aref01}
\begin{equation}
\label{Polynoials_at_positions_of_vortices}P(z)=\prod_{i=1}^{l_1}(z-a_i),\quad Q(z)=\prod_{j=1}^{l_2}(z-b_j).
\end{equation}
Note that polynomials $P(z)$ and $Q(z)$ do not have multiple and common roots. From the equations \eqref{Motion_of_Vortices} we find
\begin{equation}
\begin{gathered}
\label{Relations_for_positions_of_vortices}\sum_{i=1}^{l_1}{}^{'}\frac{1}{a_k-a_i}=\mu\sum_{j=1}^{l_2}\frac{1}{a_k-b_j},\quad k=1,\ldots, l_1,\hfill\\
\sum_{i=1}^{l_1}\frac{1}{b_m-a_i}=\mu\sum_{j=1}^{l_2}{}^{'}\frac{1}{b_m-b_j},\quad m=1,\ldots, l_2.\hfill
\end{gathered}
\end{equation}
Using properties of the logarithmic derivative, we get
\begin{equation}
\label{Polynoials_at_positions_of_vortices_Derivatives1}P_z=P\sum_{i=1}^{l_1}\frac{1}{z-a_i},\quad Q_z=Q\sum_{j=1}^{l_2}\frac{1}{z-b_j},
\end{equation}
\begin{equation}
\label{Polynoials_at_positions_of_vortices_Derivatives2}P_{zz}=P\sum_{i=1}^{l_1}\sum_{k=1}^{l_1}{}^{'}\frac{1}{(z-a_i)(z-a_k)},\quad Q_{zz}=Q\sum_{j=1}^{l_2}\sum_{m=1}^{l_2}{}^{'}\frac{1}{(z-b_j)(z-b_m)}.
\end{equation}
Equalities \eqref{Polynoials_at_positions_of_vortices_Derivatives2} can be rewritten in the form
\begin{equation}
\label{Polynoials_at_positions_of_vortices_Derivatives2_new}P_{zz}=2P\sum_{i=1}^{l_1}\sum_{k=1}^{l_1}{}^{'}\frac{1}{(z-a_i)(a_i-a_k)},\, Q_{zz}=2Q\sum_{j=1}^{l_2}\sum_{m=1}^{l_2}{}^{'}\frac{1}{(z-b_j)(b_j-b_m)}.
\end{equation}
Now let $z$ tend to one of the roots of the corresponding polynomial. Calculating the limit $z \rightarrow a_{i_0}$ in the expression for $P_{zz}$
and the limit $z\rightarrow b_{j_0}$ in the expression for $Q_{zz}$, we obtain \cite{Aref01}
\begin{equation}
\label{Derivatives2_a_j}P_{zz}(a_{i_0})=2P_z(a_{i_0})\sum_{i=1}^{l_1}{}^{'}\frac{1}{a_{i_0}-a_i},\, Q_{zz}(b_{j_0})=2Q_z(b_{j_0})\sum_{j=1}^{l_2}{}^{'}\frac{1}{b_{j_0}-b_j}.
\end{equation}
Using expressions \eqref{Relations_for_positions_of_vortices}, \eqref{Polynoials_at_positions_of_vortices_Derivatives1}, we get the conditions
\begin{equation}
\label{Conditions_at_roots}P_{zz}(a_{i_0})Q(a_{i_0})=2\mu P_z(a_{i_0})Q_z(a_{i_0}),\, Q_{zz}(b_{j_0})P(b_{j_0})=\frac{2}{\mu}Q_z(b_{j_0})P_z(b_{j_0}),
\end{equation}
which are valid for any root $a_{i_0}$ ($b_{j_0}$) of the polynomial $Q(z)$ ($P(z)$). Further we see that the polynomial $P_{zz}Q-2\mu P_zQ_z+\mu^2PQ_{zz}$ being of degree $l_1+l_2-2$ possesses $l_1+l_2$ roots $a_1$, $\ldots$, $a_{l_1}$, $b_1$, $\ldots$, $b_{l_2}$. Thus, this polynomial identically equals zero. Consequently, the generating polynomials $P(z)$ and $Q(z)$ of the arrangements  described above satisfy the differential correlation
\begin{equation}
\label{Correlation_for_Polynomials_mu}P_{zz}Q-2\mu P_zQ_z+\mu^2PQ_{zz}=0.
\end{equation}
Note that the reverse result is also valid. If two polynomials with no common and multiple roots satisfy the correlation \eqref{Correlation_for_Polynomials_mu}, then the roots of these polynomials give positions of vortices with circulations $\Gamma$ and $-\mu \Gamma$ in stationary equilibrium.

\begin{table}[b]
    \caption{Polynomials $\{P_k(z)\}$.} \label{t:P}
       \begin{tabular}[pos]{l}
                \hline
                $P_0(z) = 1$\\
        $P_1(z) = z$\\
        $P_2(z) = z^5+s_2$\\
$P_3(z) = {z}^{8}+{\frac {28}{5}}\,t_{{2}}{z}^{6}+14\,t_2^{2}{z}^{4}+28\,
t_2^{3}{z}^{2}+s_{{3}}z-7\,t_2^{4}$\\
$P_4(z) = {z}^{16}+{\frac {44}{7}}\,t_{{3}}{z}^{12}-32\,s_{{2}}{z}^{11}+22\,{t_{
{3}}}^{2}{z}^{8}-{\frac {2112}{7}}\,s_{{2}}t_{{3}}{z}^{7}+1408\,{s_{{2
}}}^{2}{z}^{6}+s_{{4}}{z}^{5}$\\
$\qquad \quad-44\,{t_{{3}}}^{3}{z}^{4}+352\,s_{{2}}{t_
{{3}}}^{2}{z}^{3}-1408\,{s_{{2}}}^{2}t_{{3}}{z}^{2}+2816\,{s_{{2}}}^{3
}z+s_{{2}}s_{{4}}-{\frac {11}{5}}\,{t_{{3}}}^{4}$\\
\hline
        \end{tabular}
\end{table}

Secondly, we consider the situation with $l_1$ vortices of circulation $\Gamma$ at positions $a_1$, $\ldots$, $a_{l_1}$,
$l_2$ vortices of circulation $-\mu \Gamma$ at positions $b_1$, $\ldots$, $b_{l_2}$, and a vortex of circulation  $-\nu \Gamma$ at
position $z=0$. Again we introduce the polynomial $P(z)$ with roots $a_1$, $\ldots$, $a_{l_1}$ and the polynomial $Q(z)$ with roots $b_1$, $\ldots$, $b_{l_2}$ (see \eqref{Polynoials_at_positions_of_vortices}). These polynomials do not have multiple and common roots. In addition we suppose that the polynomials $P(z)$ and $Q(z)$ do not have a zero root. Now, instead of correlations \eqref{Relations_for_positions_of_vortices} we have the following algebraic system
\begin{equation}
\begin{gathered}
\label{Relations_for_positions_of_vortices_new1}\sum_{i=1}^{l_1}{}^{'}\frac{1}{a_k-a_i}=\mu\sum_{j=1}^{l_2}\frac{1}{a_k-b_j}+\frac{\nu}{a_k},\quad k=1,\ldots, l_1,\hfill\\
\sum_{i=1}^{l_1}\frac{1}{b_m-a_i}=\mu\sum_{j=1}^{l_2}{}^{'}\frac{1}{b_m-b_j}+\frac{\nu}{b_m},\quad m=1,\ldots, l_2.\hfill
\end{gathered}
\end{equation}
along with the condition
\begin{equation}
\begin{gathered}
\label{Relations_for_positions_of_vortices_new2}\sum_{i=1}^{l_1}\frac{1}{a_i}=\mu\sum_{j=1}^{l_2}\frac{1}{b_j}.
\end{gathered}
\end{equation}
By analogy with the previous case we substitute expressions \eqref{Relations_for_positions_of_vortices_new1}, \eqref{Polynoials_at_positions_of_vortices_Derivatives1} into equalities \eqref{Derivatives2_a_j} and obtain
\begin{equation}
\begin{gathered}
\label{Conditions_at_roots_new}a_{i_0}P_{zz}(a_{i_0})Q(a_{i_0})=2\mu a_{i_0}P_z(a_{i_0})Q_z(a_{i_0})+2\nu P_z(a_{i_0})Q(a_{i_0}),\\
b_{j_0}Q_{zz}(b_{j_0})P(b_{j_0})=\frac{2}{\mu}b_{j_0}Q_z(b_{j_0})P_z(b_{j_0})-\frac{2\nu}{\mu}Q_z(b_{j_0})P(b_{j_0}).
\end{gathered}
\end{equation}
Now let us take the polynomial $z(P_{zz}Q-2\mu P_zQ_z+\mu^2PQ_{zz})-2\nu (P_zQ-\mu PQ_z)$, which is of degree $l_1+l_2-1$ and possesses $l_1+l_2$ roots $a_1$, $\ldots$, $a_{l_1}$, $b_1$, $\ldots$, $b_{l_2}$.
Thus we conclude that this polynomial identically equals zero
\begin{equation}
\label{Correlation_for_Polynomials_mu_new}z(P_{zz}Q-2\mu P_zQ_z+\mu^2PQ_{zz})-2\nu (P_zQ-\mu PQ_z)=0.
\end{equation}
Consequently, the vortices in arrangements described above are in stationary equilibrium if and only if the generating polynomials of the vortices satisfy differential correlation \eqref{Correlation_for_Polynomials_mu_new}. Note that from correlation \eqref{Correlation_for_Polynomials_mu_new} it follows
\begin{equation}
\label{Correlation_for_Polynomials_mu_new2}P_z(0)Q(0)-\mu P(0)Q_z(0)=0.
\end{equation}
The same result can be obtained if we use equalities \eqref{Relations_for_positions_of_vortices_new2}, \eqref{Polynoials_at_positions_of_vortices_Derivatives1}.

\begin{table}[b]
    \caption{Polynomials $\{Q_k(z)\}$.} \label{t:Q}
       \begin{tabular}[pos]{l}
                \hline
                $Q_0(z) = 1$\\
$Q_1(z) = z$\\
$Q_2(z) =z^2+t_2$\\
$Q_3(z)=z^5+t_3z-4s_2$\\
$Q_4(z)={z}^{7}+7\,t_{{2}}{z}^{5}+35\,{t_{{2}}}^{2}{z}^{3}+t_{{4}}{z}^{2}-35\,
{t_{{2}}}^{3}z-\frac52\,s_{{3}}+t_{{4}}t_{{2}}$\\
$Q_5(z)={z}^{12}+11\,t_{{3}}{z}^{8}-88\,s_{{2}}{z}^{7}+t_{{5}}{z}^{5}-77\,{t_{
{3}}}^{2}{z}^{4}+616\,s_{{2}}t_{{3}}{z}^{3}-2464\,{s_{{2}}}^{2}{z}^{2}
$\\
$\qquad \quad+t_{{5}}t_{{3}}z-\frac74\,s_{{4}}z-4\,t_{{5}}s_{{2}}+{\frac {77}{5}}\,{t_{
{3}}}^{3}$\\
\hline
        \end{tabular}
\end{table}

Now let us study polynomial solutions of equation \eqref{Correlation_for_Polynomials_mu}. If we fix, for example, the polynomial $Q(z)$, i.~e. $Q(z)=Q_k(z)$, then this relation can be regarded as  a linear second--order differential equation for the function $P(z)$. We denote by $P_{k-1}(z)$, $P_{k+1}(z)$ two linearly independent solutions of this equation with $Q(z)=Q_k(z)$. Changing the roles of $Q(z)$ and $P(z)$, we obtain the system
\begin{equation}
\begin{gathered}
\label{Correlation_for_Polynomials_System}P_{k\pm1,zz}Q_k-2\mu P_{k\pm1,z}Q_{k,z}+\mu^2P_{k\pm1}Q_{k,zz}=0,\hfill \\
P_{k,zz}Q_{k\pm1}-2\mu P_{k,z}Q_{k\pm1,z}+\mu^2P_{k}Q_{k\pm1,zz}=0. \hfill
\end{gathered}
\end{equation}
The first equation in \eqref{Correlation_for_Polynomials_System} is satisfied by the polynomials $P_{-1}(z)=1$, $Q_{0}(z)=1$, $P_{1}(z)=z$ and the second equation in its turn is satisfied by $Q_{-1}(z)=1$, $P_{0}(z)=1$, $Q_{1}(z)=z$. So we take the index $k$ in expressions \eqref{Correlation_for_Polynomials_System} as $k\in \mathbb{N}$. Using the formula relating two linearly independent solutions of a linear second--order ordinary differential equation, we obtain
\begin{equation}
\begin{gathered}
\label{Vortices_Polynomials_Solution_mu}P_{k+1}(z)=P_{k-1}(z)\int\gamma_{k+1}\frac{Q_k^{2\mu}(z)}{P_{k-1}^{2}(z)}dz,\quad k\in \mathbb{N} \cup\{0\}\hfill \\
Q_{k+1}(z)=Q_{k-1}(z)\int\delta_{k+1}\frac{P_k^{2/\mu}(z)}{Q_{k-1}^{2}(z)}dz,\quad k\in \mathbb{N} \cup\{0\}. \hfill
\end{gathered}
\end{equation}
Calculating the indefinite integrals in these expressions, we introduce the integration constant $s_{k+1}$ in the first one and $t_{k+1}$ in the second.
These formulae with arbitrary constants $s_{k+1}$, $\gamma_{k+1}$ and $t_{k+1}$, $\delta_{k+1}$ accordingly give the general solutions of equations \eqref{Correlation_for_Polynomials_System}. Note that relations \eqref{Vortices_Polynomials_Solution_mu} can be rewritten in the differential--difference way
 \begin{equation}
\begin{gathered}
\label{Vortices_Polynomials_DD_rel}P_{k+1,z}(z)P_{k-1}(z)-P_{k+1}(z)P_{k-1,z}(z)=\gamma_{k+1}Q_k^{2\mu}(z),\hfill \\
Q_{k+1,z}(z)Q_{k-1}(z)-Q_{k+1}(z)Q_{k-1,z}(z)=\delta_{k+1}P_k^{2/\mu}(z). \hfill
\end{gathered}
\end{equation}

 For the rationality of both integrals in \eqref{Vortices_Polynomials_Solution_mu} we essentially need to set $\mu=1$ , $\mu=2$, $\mu=1/2$. Now let us prove that relations \eqref{Vortices_Polynomials_Solution_mu} define polynomials. This statement is true for $k=0$. System \eqref{Correlation_for_Polynomials_System} is invariant under the transformation $z\mapsto z-z_0$. Let us set $s_1=0$, $t_1=0$. Suppose that relations \eqref{Vortices_Polynomials_Solution_mu} define polynomials at $k=1$, $\ldots$, $l-1$. Let us consider the case $k=l$. Note that at each step $k=1$, $\ldots$, $l-1$ we may define the constants $s_{k+1}$, $t_{k+1}$ in such a way that the polynomials $P_{k+1}(z)$, $Q_k(z)$, as well as the polynomials $Q_{k+1}(z)$, $P_k(z)$ do not have common roots. This fact can be proved by induction using asymptotic analysis of the integrals around a root $z=b$ of $Q_k(z)$ and around a root $z=a$ of $P_k(z)$. Besides that, the absence of common roots of the polynomials $P_{k+1}(z)$, $Q_k(z)$ and $Q_{k+1}(z)$, $P_k(z)$ implies that polynomials $P_{k+1}(z)$, $Q_{k+1}(z)$ do not have multiple roots. Indeed, assuming the contrary we come to the contradiction with the help of correlations \eqref{Vortices_Polynomials_DD_rel}. Now let us show that logarithmic terms do not appear in \eqref{Vortices_Polynomials_Solution_mu}. For this purpose it is sufficient to prove that the following conditions hold
\begin{equation}
\begin{gathered}
\label{Vortices_Residues}\text{res}_{z=a} \frac{Q_l^{2\mu}(z)}{P_{l-1}^{2}(z)}=0,\quad
\text{res}_{z=b}\frac{P_l^{2/\mu}(z)}{Q_{l-1}^{2}(z)}=0 \hfill
\end{gathered}
\end{equation}
for any simple root $z=a$ of $P_{l-1}(z)$ and any simple root $z=b$ of $Q_{l-1}(z)$. Calculating the residues, we obtain
\begin{equation}
\begin{gathered}
\label{Vortices_Residues_result}\text{res}_{z=a} \frac{Q_l^{2\mu}(z)}{P_{l-1}^{2}(z)}=\frac{Q^{2\mu-1}_l(a)}{P_{l-1,z}^3(a)}\left\{2\mu Q_{l,z}(a)P_{l-1,z}(a)-Q_l(a)P_{l-1,zz}(a)\right\},\hfill\\
\text{res}_{z=b}\frac{P_l^{2/\mu}(z)}{Q_{l-1}^{2}(z)}=\frac{P^{2/\mu-1}_l(b)}{Q_{l-1,z}^3(b)}\left\{\frac{2}{\mu} P_{l,z}(b)Q_{l-1,z}(b)-P_l(b)Q_{l-1,zz}(b)\right\} \hfill
\end{gathered}
\end{equation}
Using equalities \eqref{Correlation_for_Polynomials_System}, we see that conditions \eqref{Vortices_Residues} are valid. Thus formulae \eqref{Vortices_Polynomials_Solution_mu} enables one to construct sequences of polynomials satisfying equation \eqref{Correlation_for_Polynomials_mu} essentially in the following cases $\mu=1$ , $\mu=2$, $\mu=1/2$. The cases $\mu=2$, $\mu=1/2$ are equivalent since we may change the roles of $Q(z)$ and $P(z)$. The sequence of polynomials, which provide solutions of equation \eqref{Correlation_for_Polynomials_mu} with $\mu=1$ is  given by the Adler -- Moser polynomials~\cite{Adler01}. Note that in this case we set $P_k(z)=Q_k(z)$. The Adler -- Moser polynomials are associated with rational solutions of the Korteweg -- de Vries equation. The case $\mu=2$ in relationship with the equilibrium of charges in the plane was studied in the article \cite{Loutsenko01}. We would like to emphasize that polynomials from the sequences $\{P_k(z)\}$, $\{Q_k(z)\}$ may have multiple or common roots at certain values of the parameters $\{s_k\}$, $\{t_k\}$. Along with this there may exist polynomial solutions of equation \eqref{Correlation_for_Polynomials_mu} with $\mu=1$, $\mu=2$ that are can not be included into the sequences $\{P_k(z)\}$, $\{Q_k(z)\}$.

In what follows we shall consider the case $\mu=2$. The degrees of the polynomials can be obtained balancing the highest--order terms in relation \eqref{Correlation_for_Polynomials_mu}. Consequently, we find
\begin{equation}
\begin{gathered}
\label{Vortices_Polynomials_degrees}\deg P_k(z)=\frac{6k(k+1)-1+(-1)^k(2k+1)}{8},\quad \\
\\
\deg Q_k(z)=\frac{6k(k+1)+1+(-1)^{k+1}(2k+1)}{16}.
\end{gathered}
\end{equation}
Let us take the parameters $\gamma_{k+1}$, $\delta_{k+1}$ in expressions \eqref{Vortices_Polynomials_Solution_mu} in such a way that all the polynomials are monic, thus we get
\begin{equation}
\begin{gathered}
\label{Vortices_Polynomials_degrees2}\gamma_{k+1}=\frac{6k+3+(-1)^{k+1}}{2},\quad \delta_{k+1}=\frac{6k+3+(-1)^{k}}{4}.
\end{gathered}
\end{equation}
Thus we have proved that roots of the polynomials $P_k(z)$, $Q_{k+1}(z)$ ($P_{k+1}(z)$, $Q_{k}(z)$) give positions of vortices with circulations $\Gamma$ (roots of $P_{k}(z)$ or $P_{k+1}(z)$) and $-2\Gamma$ (roots of $Q_{k+1}(z)$ or $Q_{k}(z)$) in stationary equilibrium whenever the polynomials $P_k(z)$, $Q_{k+1}(z)$ ($P_{k+1}(z)$, $Q_{k}(z)$) do not have multiple and common roots.

\begin{figure}[t]
 \centerline{
 \subfigure[$Q_4(z)$]{\epsfig{file=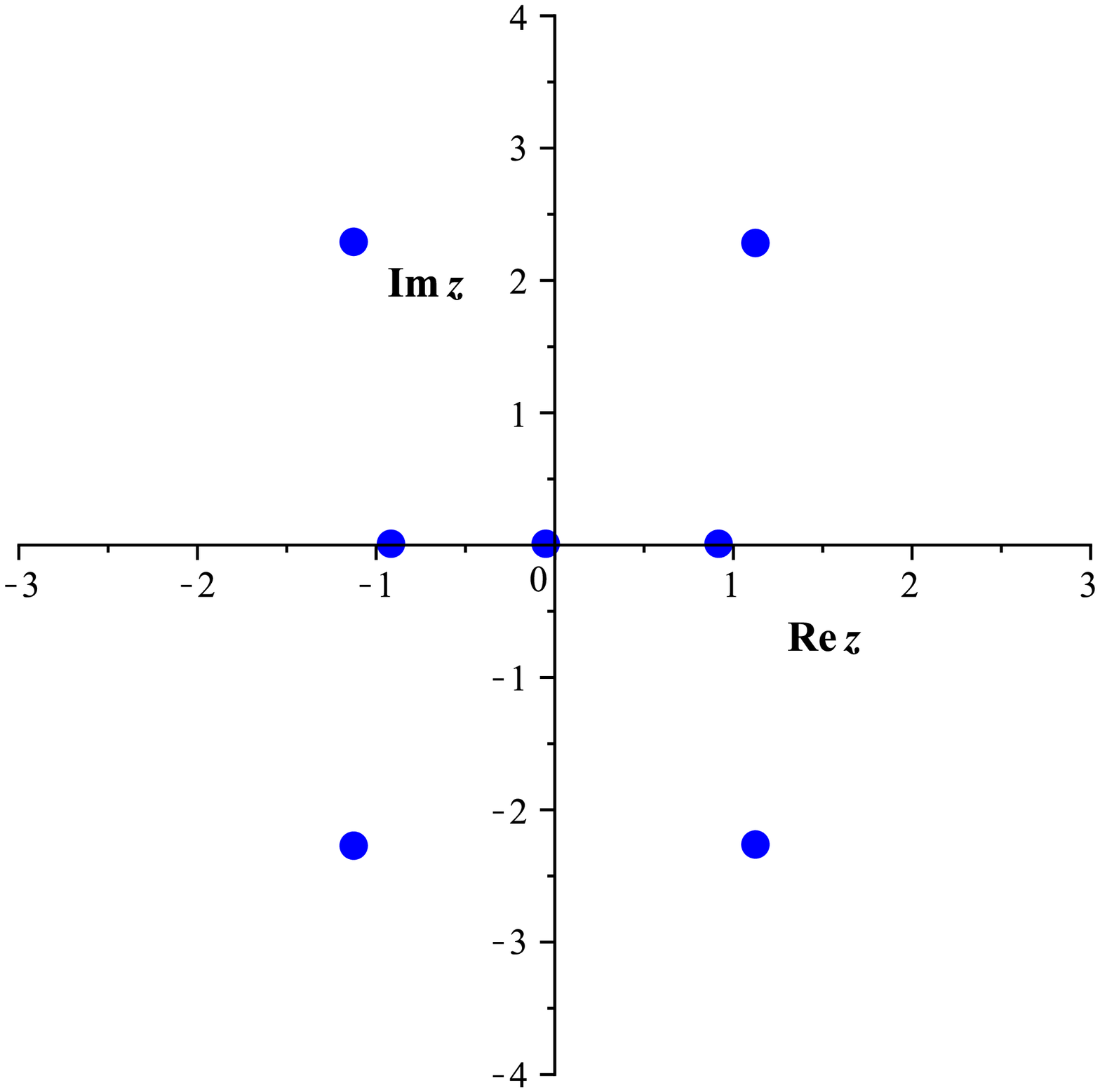,width=60mm}}
 \subfigure[$P_5(z)$]{\epsfig{file=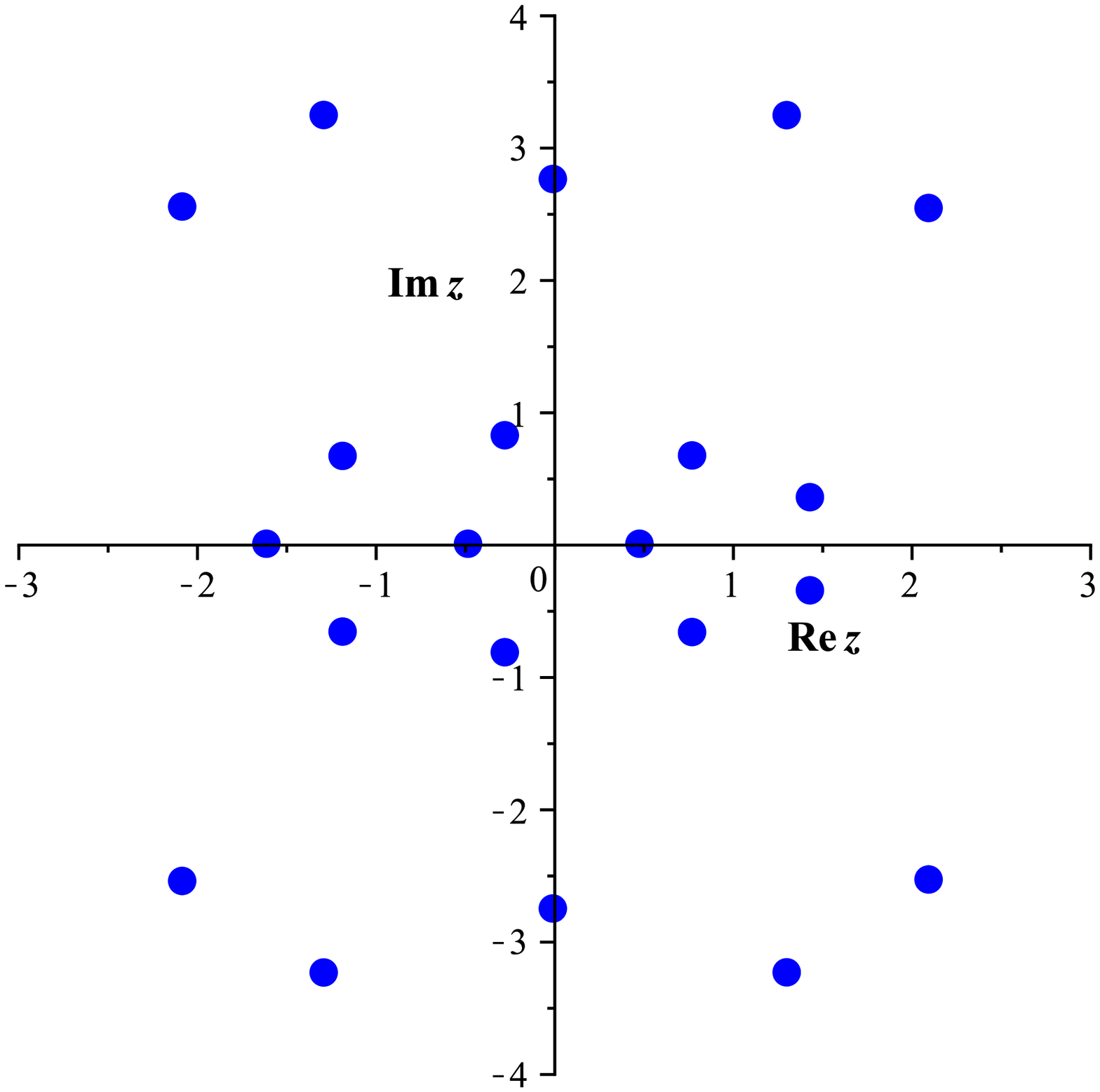,width=60mm}}}
   \caption{Roots of the polynomials $\{Q_k(z)\}$, $\{P_k(z)\}$, $t_2=1$, $s_3=1$, $t_4=1$, $s_5=1$.}
 \label{F:Plots1}
\end{figure}

\begin{figure}[t]
 \centerline{
   \subfigure[$Q_5(z)$]{\epsfig{file=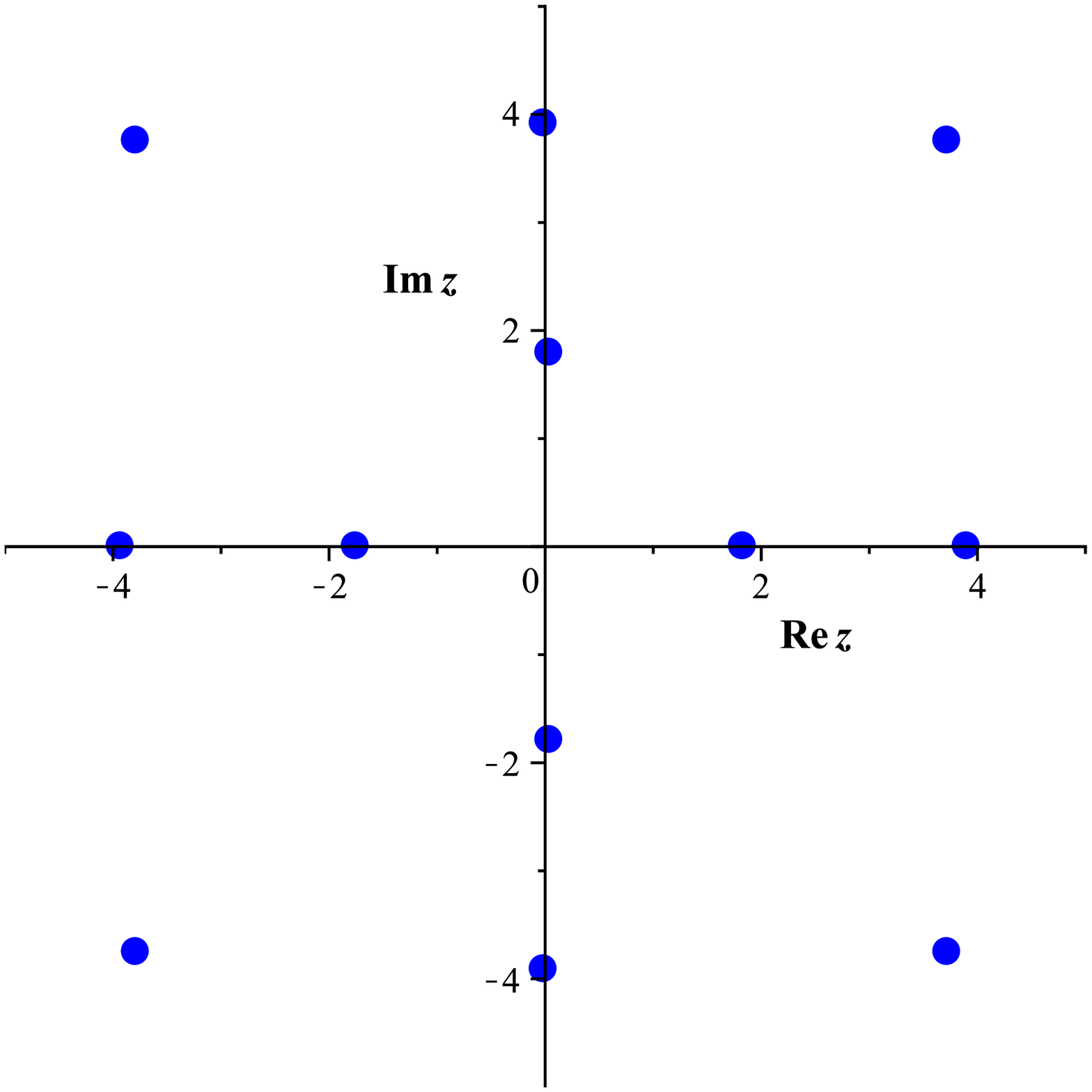,width=60mm}}
 \subfigure[$P_4(z)$]{\epsfig{file=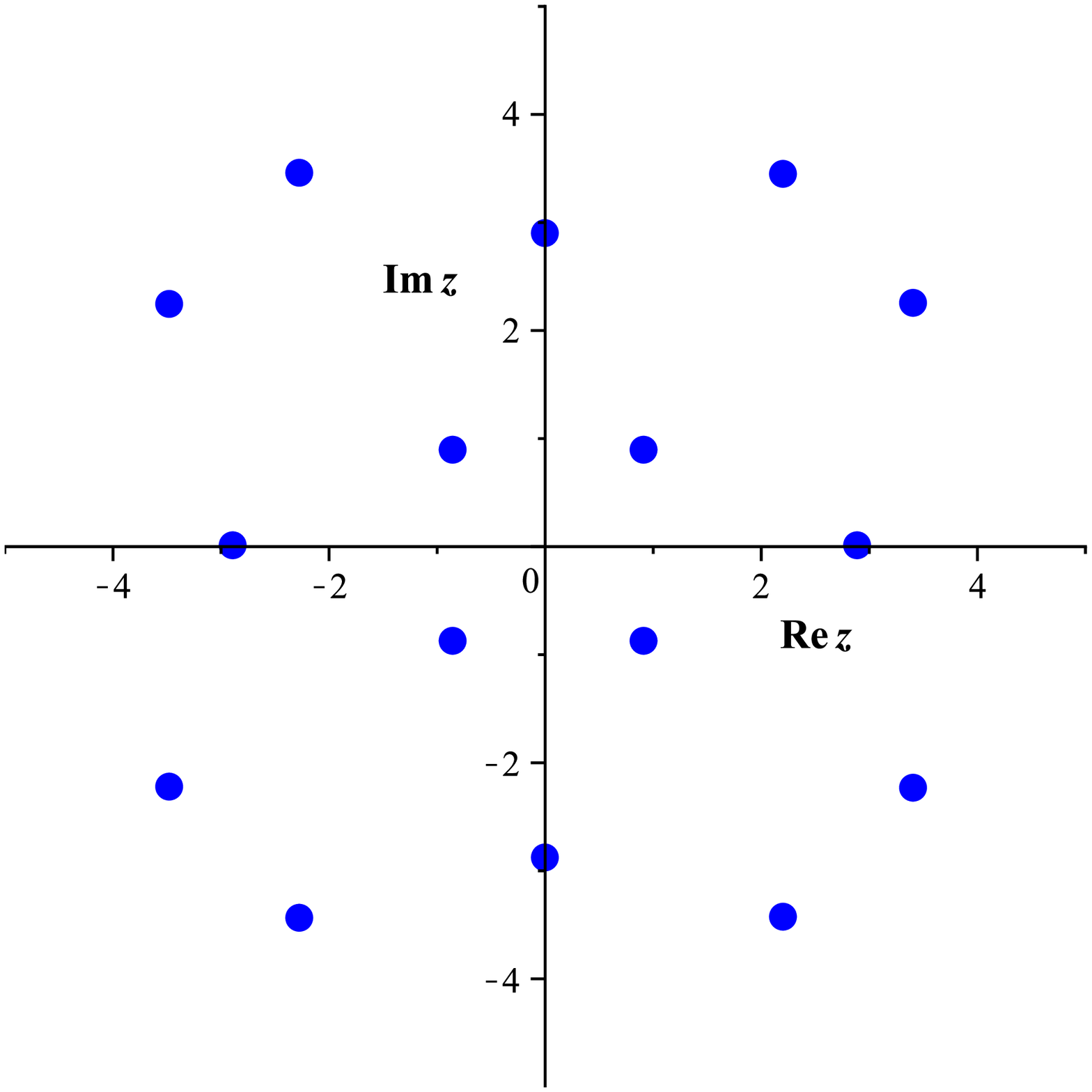,width=60mm}}}
 \caption{Roots of the polynomials $\{Q_k(z)\}$, $\{P_k(z)\}$, $s_2=1$, $t_3=50$, $s_4=1$, $t_5=1$.}
 \label{F:Plots2}
\end{figure}

First few polynomials $\{P_k(z)\}$, $\{Q_k(z)\}$ in explicit form are given in tables \ref{t:P}, \ref{t:Q}. Interestingly, that roots of the polynomials in question form regular structures in the complex plane. For example, see figures \ref{F:Plots1}, \ref{F:Plots2}.

In next sections we shall study special polynomials associated with rational solutions of the generalized $K_2$ hierarchy. Our aim is to show that partial cases of the polynomials $\{P_k(z)\}$, $\{Q_k(z)\}$ appear in the theory of integrable differential equations.

\section{Rational solutions of the generalized $K_2$ hierarchy} \label{Gen_K2}

The  Sawada -- Kotera equation, one of the partial differential equations integrable by the inverse scattering method,
\begin{equation}
\label{SK_Eq}u_t+u_{xxxxx}-5uu_{xxx}-5u_xu_{xx}+5u^2u_x=0
\end{equation}
possesses the self--similar reduction $u(x,t)=(5\,t)^{-\frac25}\,\left\{w_z(z)+w^2(z)\right\}$, $z=x\,(5\,t)^{-\frac15}$ with $w(z)$ satisfying the equation
\begin{equation}
\label{1.1}
w_{zzzz}+5\,w_{z}\,w_{zz}-5\,w^2\,w_{zz}-5\,w\,w_{z}^{2}+w^5-z\,w-\beta=0.
\end{equation}
Another soliton equation, the Kaup -- Kupershmidt equation
\begin{equation}
\label{KK_Eq}v_t+v_{xxxxx}+10vv_{xxx}+25v_xv_{xx}+20v^2v_x=0
\end{equation}
also admits a self--similar reduction to equation \eqref{1.1}. Indeed setting $v(x,t)=(5\,t)^{-\frac25}\,\left\{w_z(z)-\frac{w^2(z)}{2}\right\}$, $z=x\,(5\,t)^{-\frac15}$, we obtain equation \eqref{1.1} for the function $w(z)$.

  In this section we consider the following hierarchy of ordinary differential equations \cite{Kudr99}
\begin{equation}\begin{gathered}
\label{ku0.1}
K_2^{(N)}[\beta_N,\alpha_1,\ldots,\alpha_N]:\quad \left(\frac{d}{dz}+w\right)\sum_{n=1}^{N}\alpha_nH_{n}\left[w_z-\frac12\,w^2\right]-\\
\\
-z\,w-\beta_N=0,\quad N=1, 2, \ldots,
\end{gathered}
\end{equation}
where the function $H_{n}[v]$ is defined through the recurrence formula
\begin{equation}\begin{gathered}
\label{ku0.2}H_{n+2}[v]=J_1[v]\,\Omega[v]\,H_{n}[v],\quad H_{0}[v]=1,\quad H_{1}[v]=v_{zz}+4\,v^2
\end{gathered}\end{equation}
and the operators $J_1[v]$ and $\Omega[v]$ are given by
\begin{equation}\begin{gathered}
\label{ku0.4}\Omega[v]=D^3+2\,v\,D+v_z,\,\,\quad\,D=\frac{d}{dz}
\end{gathered}\end{equation}
\begin{equation}\begin{gathered}
\label{ku0.5}J_1[v]=D^3+3\left(vD+D\,v\right)+2\,\left(D^2\,v\,D^{-1}+D^{-1}\,v\,D^2\right)
+\\
\\
+8 \left(v^2\,D^{-1}+D^{-1}\,v^2\right),\,\quad\, D^{-1}=\int\,dz
\end{gathered}\end{equation}
The hierarchy \eqref{ku0.1} arises as a self--similar reduction of the Sawada -- Kotera equation \eqref{SK_Eq}, the Kaup -- Kupershmidt equation \eqref{KK_Eq}, and their hierarchies. Let us call the hierarchy \eqref{ku0.1} as the generalized $K_2$ hierarchy. Note that the equations in the generalized $K_2$ hierarchy admit another representation
\begin{equation}\begin{gathered}
\label{ku0.6}
K_2^{(N)}[\beta_N,\alpha_1,\ldots,\alpha_N]:\quad -\frac12\left(\frac{d}{dz}-2\,w\right)\sum_{n=1}^{N}\alpha_nG_{n}\left[-2\,w_z-
2\,w^2\right]-\\
\\
-z\,w-\beta=0,\quad N=1, 2, \ldots,
\end{gathered}
\end{equation}
where the function $G_{n}[u]$ is defined as
\begin{equation}
\begin{gathered}
\label{ku0.7}G_{n+2}[u]=J_{2}[u]\,\Omega[u]\,G_{n}[u],\quad G_{0}[u]=1,\quad G_{1}[u]=u_{zz}+\frac{1}{4}\,u^2
\end{gathered}
\end{equation}
and the operator $J_2[u]$ takes the form
\begin{equation}\begin{gathered}
\label{ku0.9}J_{2}[u]=D^3+\frac{1}{2}\left(D^2\,u\,D^{-1}+D^{-1}\,u\,D^2\right)+ \frac{1}{8} \left(u^2\,D^{-1}+D^{-1}\,u^2\right).
\end{gathered}\end{equation}
In what follows we introduce notation
\begin{equation}
\begin{gathered}
\label{New_operators} \tilde{H}_{N}[v]=\sum_{n=1}^{N}\alpha_nH_{n}[v],\quad \displaystyle \tilde{G}_{N}[u]=\sum_{n=1}^{N}\alpha_nG_{n}[u].
\end{gathered}
\end{equation}
The parameter $\alpha_N$ can be removed from \eqref{ku0.1} and \eqref{ku0.6} by means of a simple transformation, but for convenience we shall not do this.

It is a distinguishing property of the Painlev\'{e} equations and higher--order analogues of the Painlev\'{e} equations that their rational solutions, if any, admit representation
in terms of certain special polynomials \cite{Clarkson01,Kudr99a, Kudr02}. Rational solutions of the equations in the generalized $K_2$  hierarchy also admit representation via logarithmic derivative of special polynomials. The rational solutions are classified in the following theorem.

\textbf{Theorem 1.}
The equation $K_2^{(N)}[\beta_N,\alpha_1,\ldots,\alpha_N]$ possesses rational solutions if and only if $\beta_N$ $\in$ $\mathbb{Z}\setminus\{1\pm 3n, n\in\mathbb{N}\}$. They are unique at fixed values of the parameters $\alpha_1$, $\ldots$, $\alpha_N$ and have the form
\begin{equation}\begin{gathered}
\label{K2_RS}w^{(N)}(z;\beta_k^{(1)})=\frac{d}{d z}\ln \frac{S_{k-1}}{T_k^2},\quad w^{(N)}(z;\beta_k^{(2)})=\frac{d}{d z}\ln \frac{S_{k}}{T_{k-1}^2},\quad k\in\mathbb{N}
\end{gathered}
\end{equation}
where $S_k(z)\equiv S_k^{(N)}(z)$, $T_k(z)\equiv T_k^{(N)}(z)$ are polynomials of degrees
\begin{equation}\begin{gathered}
\label{TS_degrees}\deg S_k(z)=\frac{6k(k+1)-1+(-1)^k(2k+1)}{8},\\
\\
\quad \deg T_k(z)=\frac{6k(k+1)+1+(-1)^{k+1}(2k+1)}{16}
\end{gathered}
\end{equation}
and the parameter $\beta$ is given by
\begin{equation}\begin{gathered}
\label{RS_beta}\beta_k^{(1)}=\frac32 k+\frac14\left\{1+(-1)^{k+1}\right\},\quad \beta_k^{(2)}=-\frac32 k+\frac14\left\{1+(-1)^{k+1}\right\}.
\end{gathered}
\end{equation}
The only remaining rational solution is the trivial solution $w^{(N)}(z;0)=0$.

This theorem can be proved by analogy with the case of the generalized second Painlev\'{e} hierarchy performing asymptotic analysis for solutions of the equation $K_2^{(N)}[\beta_N,\alpha_1,\ldots,\alpha_N]$ (for more details see, for example \cite{Kudr10, Demina15, Kudr08, Kudr07}). Note that the sequences of polynomials begin with $S_0^{(N)}(z)=1$, $S_1^{(N)}(z)=z$, $T_0^{(N)}(z)=1$, $T_1^{(N)}(z)=z$. The rational solutions of the equation $K_2^{(N)}[\beta_N,\alpha_1,\ldots,\alpha_N]$ can be calculated in explicit form with the help of B\"{a}cklund transformations. For any solution $w\equiv w^{(N)}(z,\beta_N)$ of the equation $K_2^{(N)}[\beta_N,\alpha_1,\ldots,\alpha_N]$ the following transformations
\begin{equation}\begin{gathered}
\label{K2_BT}w^{(N)}(z,2-\beta_N)=w-\frac{2\,\beta_N-2}{\tilde{H}_{N}
\left[w_z-\frac12\,w^2\right]-z},\hfill \\
w^{(N)}(z,-1-\beta_N)=w-\frac{2\,\beta_N+1}{\tilde{G}_{N}
\left[-2\,w_z-2\,w^2\right]-z},\hfill
\end{gathered}\end{equation}
produce solutions of the same equation with different values of the parameter $\beta_N$.

Further let us derive several differential -- difference relations satisfied by the polynomials $S_k^{(N)}(z)$, $T_k^{(N)}(z)$.

\textbf{Theorem 2.}
The polynomials
$S_k(z)\equiv S_k^{(N)}(z)$, $T_k(z)\equiv T_k^{(N)}(z)$ satisfy the following
differential -- difference relations
\begin{equation}
\begin{gathered}
\label{ST_Rec_rel1}S_{k\pm1,zz}T_k-4 S_{k\pm1,z}T_{k,z}+4S_{k\pm1}T_{k,zz}=0,\hfill \\
S_{k,zz}T_{k\pm1}-4 S_{k,z}T_{k\pm1,z}+4S_{k}T_{k\pm1,zz}=0. \hfill
\end{gathered}
\end{equation}
and
\begin{equation}
\begin{gathered}
\label{ST_Rec_rel2}S_{k+1,z}S_{k-1}-S_{k+1}S_{k-1,z}=\left(3k+\frac{(-1)^{k+1}}{2}+\frac32\right)T_k^4,\hfill \\
T_{k+1,z}T_{k-1}-T_{k+1}T_{k-1,z}=\left(\frac{3k}{2}+\frac{(-1)^{k}}{4}+\frac34\right)S_k. \hfill
\end{gathered}
\end{equation}

\textbf{Proof.}
Without loss of generality, we fix the $N^{\text{th}}$ equation in the generalized $K_2$ hierarchy and omit the index $N$. By direct substitution of the B\"{a}cklund transformations \eqref{K2_BT}, we can prove that the following formulae are valid
\begin{equation}
\begin{gathered}
\label{ST_Invariants}\frac{d}{dz}w\left(z;\beta_{k+1}^{(1)}\right)-\frac{1}{2}w^2\left(z;\beta_{k+1}^{(1)}\right)= \frac{d}{dz}w\left(z;\beta_{k}^{(2)}\right)-\frac{1}{2}w^2\left(z;\beta_{k}^{(2)}\right),\quad \beta_{k+1}^{(1)}=2-\beta_{k}^{(2)},\hfill\\
\frac{d}{dz}w\left(z;\beta_{k}^{(1)}\right)+w^2\left(z;\beta_{k}^{(1)}\right)= \frac{d}{dz}w\left(z;\beta_{k+1}^{(2)}\right)+w^2\left(z;\beta_{k+1}^{(2)}\right),\quad \beta_{k+1}^{(2)}=-1-\beta_{k}^{(1)}. \hfill
\end{gathered}
\end{equation}
First of all, let us derive  correlations \eqref{ST_Rec_rel1}.
Substituting the representation of the rational solutions in terms of the polynomials $S_k(z)$, $T_k(z)$ into formulae \eqref{ST_Invariants}, we obtain
\begin{equation}
\begin{gathered}
\label{ST_Rec_rel_derivation}\frac{S_{k,zz}T_{k+1}-4 S_{k,z}T_{k+1,z}+4S_{k}T_{k+1,zz}}{S_kT_{k+1}}= \frac{S_{k,zz}T_{k-1}-4 S_{k,z}T_{k-1,z}+4S_{k}T_{k-1,zz}}{S_kT_{k-1}}, \hfill \\
\frac{S_{k+1,zz}T_{k}-4 S_{k+1,z}T_{k,z}+4S_{k+1}T_{k,zz}}{S_{k+1}T_{k}}= \frac{S_{k-1,zz}T_{k}-4 S_{k-1,z}T_{k,z}+4S_{k-1}T_{k,zz}}{S_{k-1}T_{k}}.\hfill
\end{gathered}
\end{equation}
Decreasing the subscript in \eqref{ST_Rec_rel_derivation} and using "initial conditions" $S_0(z)=1$, $S_1(z)=z$, $T_0(z)=1$, $T_1(z)=z$, gives required relations \eqref{ST_Rec_rel1}.

Further let us prove that differential -- difference relations \eqref{ST_Rec_rel2} hold. Again we use formulae \eqref{ST_Invariants} to get
\begin{equation}
\begin{gathered}
\label{ST_Rec_rel_derivation2}\frac{d^2}{dz^2}\ln\frac{S_{k-1}}{S_{k+1}}=
\left(\frac{d}{dz}\ln\frac{S_{k-1}}{S_{k+1}}\right)
\left(\frac{d}{dz}\ln\frac{T_k^4}{S_{k+1}S_{k-1}}\right),\hfill \\
\frac{d^2}{dz^2}\ln\frac{T_{k-1}}{T_{k+1}}=
\left(\frac{d}{dz}\ln\frac{T_{k-1}}{T_{k+1}}\right)
\left(\frac{d}{dz}\ln\frac{S_{k}^2}{T_{k+1}T_{k-1}}\right).\hfill
\end{gathered}
\end{equation}
Integrating these expressions, we obtain
\begin{equation}
\begin{gathered}
\label{ST_Rec_rel_derivation3}S_{k+1,z}(z)S_{k-1}(z)-S_{k+1}(z)S_{k-1,z}(z)=\lambda_{k+1}T_k^{4}(z),\hfill \\
T_{k+1,z}(z)T_{k-1}(z)-T_{k+1}(z)T_{k-1,z}(z)=\xi_{k+1}S_k(z). \hfill
\end{gathered}
\end{equation}
Balancing the leading order term in \eqref{ST_Rec_rel_derivation3}, we find the values of the parameters  $\lambda_{k+1}$, $\xi_{k+1}$ as given in \eqref{ST_Rec_rel2}. This completes the proof.

Thus we see that the polynomials associated with rational solutions of the equations in the generalized $K_2$ hierarchy satisfy the correlations, which we have obtained for describing equilibria of vortices with circulations $\Gamma$ and $-2\Gamma$. In fact the polynomial $S_k^{(N)}(z)$ is a partial case of the polynomial $P_k(z)$ with $\mu=2$ and the polynomial $T_k^{(N)}(z)$ is a partial case of the polynomial $Q_k(z)$ with $\mu=2$ (see the previous section). The roots of the polynomials $S_k^{(N)}(z)$, $T_{k+1}^{(N)}(z)$ or $S_{k+1}^{(N)}(z)$, $T_{k}^{(N)}(z)$ give positions of vortices with circulations $\Gamma$ and $-2\Gamma$ in stationary equilibrium, whenever the corresponding polynomials do not have multiple and common roots.

Let us derive formulae for constructing the polynomials $S_k^{(N)}(z)$, $T_k^{(N)}(z)$ in explicit form. In what follows we omit the upper index $N$. For this aim let us calculate the expressions $u=-2(w_z+w^2)$, $v=w_z-\frac12w^2$. Using the differential -- difference correlations of theorem 2 
, we obtain
\begin{equation}
\begin{gathered}
\label{ST_uv}u(z;\beta_k^{(1)})=12\frac{d^{\,2}}{dz^2}\ln T_k,\quad v(z;\beta_{k}^{(2)})=\frac32\frac{d^{\,2}}{dz^2}\ln S_k.
\end{gathered}
\end{equation}
Along with this the following relations are valid
\begin{equation}
\begin{gathered}
\label{ST_Rel_for_RF}w(z;\beta_{k+1}^{(2)})-w(z;\beta_k^{(1)})=\frac{d}{dz}\ln\frac{S_{k+1}}{S_{k-1}},\quad \beta_{k+1}^{(2)}=-1-\beta_{k}^{(1)},\hfill \\
w(z;\beta_{k+1}^{(1)})-w(z;\beta_k^{(2)})=2\frac{d}{dz}\ln\frac{T_{k-1}}{T_{k+1}},\quad \beta_{k+1}^{(1)}=2-\beta_{k}^{(2)}.\hfill
\end{gathered}
\end{equation}
Substituting these expressions and differential -- difference formulae \eqref{ST_Rec_rel2} into the B\"{a}cklund transformations \eqref{K2_BT}, we get the recurrence relations, which enable one to construct the polynomials explicitly
\begin{equation}\label{ST_Rel_Explicit}
\begin{gathered}
S_{k+1}S_{k-1}=T^4_k\left(z-\tilde{G}_N\left[12\frac{d^{\,2}}{dz^2} \ln T_k\right]\right), \\
T_{k+1}T_{k-1}=S_k\left(z-\tilde{H}_N\left[\frac32\frac{d^{\,2}}{dz^2} T_k\right]\right). \hfill
\end{gathered}
\end{equation}
Now let us derive ordinary differential equations satisfied by the polynomials $S_k^{(N)}(z)$, $T_k^{(N)}(z)$. The Miura transformation $v=-2(w_z+w^2)$ relates solutions of the equation $K_2^{(N)}[\beta_N,\alpha_1,\ldots,\alpha_N]$ to solutions of the following ordinary differential equation
\begin{equation}
\begin{gathered}
\label{SK_ordinary}\left(\frac{d^{\,3}}{dz^3}+2v\frac{d}{dz}+v_z\right)\tilde{H}_N[v]-zv_z-2v=0.
\end{gathered}
\end{equation}
Consequently, substituting the function $v$ as given in \eqref{ST_uv} into this equation, we obtain an ordinary differential equation satisfied by the polynomials $S_k^{(N)}(z)$. Similarly, for any solution $w$ of the equation $K_2^{(N)}[\beta_N]$ there exists are solution $u=w_z-\frac12 w^2$ of the following differential equation
\begin{equation}
\begin{gathered}
\label{KK_ordinary}\left(\frac{d^{\,3}}{dz^3}+2u\frac{d}{dz}+u_z\right)\tilde{G}_N[u]-zu_z-2u=0.
\end{gathered}
\end{equation}
Thus, making the substitution of the function $u$ given by \eqref{ST_uv} into the equation \eqref{KK_ordinary} yields an ordinary differential equation satisfied by the polynomials $T_k^{(N)}(z)$.

In conclusion we would like to note that the Sawada -- Kotera equation in the form
\begin{equation}
\label{SK_Eq_Alternative}5u_{\alpha_1}+u_{zzzzz}-5uu_{zzz}-5u_zu_{zz}+5u^2u_z=0,\quad u=u(\alpha_1,z)
\end{equation}
possesses rational solutions expressible in terms of the polynomials $T_k^{(N)}(z)$, i.e. $u=-6\{\ln T_k^{(N)}(z)\}_{zz}$.
Analogously, the Kaup -- Kupershmidt equation in the form
\begin{equation}
\label{KK_Eq_Alternative}5v_{\alpha_1}+v_{zzzzz}+10vv_{zzz}+25v_zv_{zz}+20v^2v_z=0,\quad v=v(\alpha_1,z)
\end{equation}
possesses rational solutions, which can be expresses via the polynomials $S_k^{(N)}(z)$. Indeed we have $v=3\{\ln S_k^{(N)}(z)\}_{zz}/2$.

\section{Rational solutions of the fourth -- order equation in the generalized $K_2$ hierarchy} \label{Gen_K2_1}

\begin{table}[b]
    \caption{Polynomials $\{S_k^{(1)}(z)\}$.} \label{t:S_K2_1}
       \begin{tabular}[pos]{l}
                \hline
                $S_0^{(1)}(z) = 1$\\
        $S_1^{(1)}(z) = z$\\
        $S_2^{(1)}(z) = z^5+36\alpha_1$\\
$S_3^{(1)}(z) = {z}^{8}$\\
$S_4^{(1)}(z) = {z}^{16}-1152\,\alpha_{{1}}{z}^{11}+1824768\,{\alpha_{{1}}}^{2}{z}^{6}
+131383296\,{\alpha_{{1}}}^{3}z$\\
$S_5^{(1)}(z) = {z}^{21}-3276\,\alpha_{{1}}{z}^{16}+6604416\,{\alpha_{{1}}}^{2}{z}^{11
}+3328625664\,{\alpha_{{1}}}^{3}{z}^{6}$\\
$\qquad \quad-119830523904\,{\alpha_{{1}}}^{
4}z$\\
$S_6^{(1)}(z) = {z}^{33}-15840\,\alpha_{{1}}{z}^{28}+63866880\,{\alpha_{{1}}}^{2}{z}^{
23}+708155965440\,{\alpha_{{1}}}^{3}{z}^{18}$ \qquad \qquad\\
$\qquad \quad+1922177762806726656\,{
\alpha_{{1}}}^{5}{z}^{8}$\\
\hline
        \end{tabular}
\end{table}

\begin{table}[b]
    \caption{Polynomials $\{T_k^{(1)}(z)\}$.} \label{t:T_K2_1}
       \begin{tabular}[pos]{l}
                \hline
                $T_0^{(1)}(z) = 1$\\
$T_1^{(1)}(z) = z$\\
$T_2^{(1)}(z) =z^2$\\
$T_3^{(1)}(z)=z^5-144\alpha_1$\\
$T_4^{(1)}(z)={z}^{7}-504\alpha_1z^2$\\
$T_5^{(1)}(z)={z}^{12}-3168\,\alpha_1\,{z}^{7}-3193344\,{\alpha_1}^{2}{z}^{2}$\\
$T_6^{(1)}(z)={z}^{15}-6552\,\alpha_1\,{z}^{10}-13208832\,{\alpha_1}^{2}{z}^{5}-
951035904\,{\alpha_1}^{3} \qquad \qquad \qquad$\\
\hline
        \end{tabular}
\end{table}

In this section we study the case of the fourth -- order equation in the generalized $K_2$ hierarchy. This equation can be written as
\begin{equation}
\label{K2_1}
\alpha_1(w_{zzzz}+5\,w_{z}\,w_{zz}-5\,w^2\,w_{zz}-5\,w\,w_{z}^{2}+w^5)-z\,w-\beta_1=0,\quad \alpha_1\neq0.
\end{equation}
Rational solutions of  equation \eqref{K2_1} are given in theorem 1  (see also \cite{Kudr07}). The special polynomials $S_k^{(1)}(z)$, $T_k^{(1)}(z)$ appearing in theorem 1 can be generated with the help of the following differential -- difference relations
\begin{equation}
\begin{gathered}
\label{ST_Rel_Explicit_1}S_{k+1}S_{k-1}=zT^4_k-12\alpha_1(T_k^3T_{k,zzzz}-4T^2_kT_{k,z}T_{k,zzz}+6T_kT_{k,z}^2T_{k,zz}-3T_{k,z}^4),\hfill\\
T_{k+1}S_kT_{k-1}=zS_k^2-\frac32\alpha_1\left(S_kS_{k,zzzz}-4S_{k,z}S_{k,zzz}+3S_{k,zz}^2\right).
\end{gathered}
\end{equation}
Here and in what follows we set $N=1$ and omit this index. Several examples of these polynomials are give in tables \ref{t:S_K2_1}, \ref{t:T_K2_1}.

Now let us derive sixth--order ordinary differential equations satisfied by the polynomials. We take equations \eqref{SK_ordinary},  \eqref{KK_ordinary}, make the substitution $v=3h_z/2$ in the first one and $u=12g_z$ in the second, and after that integrate the results to obtain
\begin{equation}
\begin{gathered}
\label{ODS_ST_1}\alpha_1\left(h_{zzzzz}+15h_zh_{zzz}+\frac{45}{4}h_{zz}^2+15h_z^3\right)-zh_z-h=0,\quad h=\left(\ln S_k\right)_{z},\hfill\\
\alpha_1\left(g_{zzzzz}+30g_zg_{zzz}+60g_z^3\right)-zg_z-g=0,\quad g=\left(\ln T_k\right)_{z}.\hfill
\end{gathered}
\end{equation}
In this expressions we set the constants of integration to zero as otherwise the equations do not have rational solutions with asymptotic behavior in a neighborhood of infinity $h(z)=\deg S_k/z$, $g(z)=\deg T_k/z$. This fact can be proved if we perform asymptotic analysis around infinity for the solutions of these equations. Now let us study properties of the polynomials $\{S_k(z)\}$, $\{T_k(z)\}$.

\textbf{Theorem 3.}
The following statements hold

\begin{enumerate}

\item The polynomials $S_k(z)$, $T_k(z)$ do not have multiple roots except possibly the point $z=0$.
\item The polynomials $S_k(z)$, $T_{k\pm1}(z)$ and $S_{k\pm1}(z)$, $T_{k}(z)$ do not have common roots except possibly the point $z=0$.
\item If the point $z=0$ is the multiple root of the polynomial $S_k$, then its multiplicity equals $8$.
\item If the point $z=0$ is the multiple root of the polynomial $T_k$, then its multiplicity equals $2$.
\item If the point $z=0$ is a root of the polynomial $S_k$ of multiplicity $8$, then the point $z=0$ is a root of the polynomials $T_{k+1}$, $T_{k-1}$ of multiplicity $2$ and a simple root of the polynomials $S_{k+2}$, $S_{k-2}$.
\item If the point $z=0$ is a root of the polynomial $T_k$ of multiplicity $2$, then the point $z=0$ is a root of the polynomials $S_{k+1}$, $S_{k-1}$ and the total multiplicity is equal to $9$.
\end{enumerate}

\textbf{Proof.}
Making asymptotic analysis around zeros for the solutions of the equations satisfied by the polynomials  $S_k$, $T_k$ (see \eqref{ODS_ST_1}), we find that the polynomial $S_k$ may have simple roots and roots of multiplicity $8$ and  the polynomial $T_k$ may have simple roots and roots of multiplicity $2$. Further constructing the Laurent series in a neighborhood of poles and infinity for solutions of the equation \eqref{K2_1} yields the structure of its rational solutions
\begin{equation}
\begin{gathered}
\label{K2_1_RS}w(z)=\sum_{i=1}^{n_{1}}\frac{1}{z-z_{1,i}}+\sum_{i=1}^{n_{4}}\frac{4}{z-z_{4,i}}-\sum_{i=1}^{n_{-2}}\frac{2}{z-z_{-2,i}}-\sum_{i=1}^{n_{-3}}\frac{3}{z-z_{-3,i}}.
\end{gathered}
\end{equation}
The rational solution \eqref{K2_1_RS} possesses $n_j$ poles with residue $j$, $j=1$, $4$, $-2$, $-3$. If we combine this expression with the formula giving representation of the rational solutions in terms of the polynomials $S_k$, $T_k$ (see theorem \ref{Teorem:RS}), we obtain the fifth statement of the theorem with the point $z=0$ replaced by the point $z=z_0$. Moreover, if a point $z=z_0$ is a multiple root of the polynomial $T_k$, then this point is also a root of the polynomials $S_{k+1}$, $S_{k-1}$ of multiplicities either $1$, or $8$. Further suppose that $z=z_0$ is a multiple root of the polynomial $T_{k}$, then the multiplicity of this root is $2$. Finding the Taylor series for the polynomial $T_k$ around its multiple root $z=z_0$ with the help of the equation for the polynomials given in \eqref{ST_Rel_Explicit_1} and  substituting this series into the right--hand side of the first expression in \eqref{ST_Rel_Explicit_1}, we see that it is satisfied only in the case $z_0=0$. Along with this, we see that the total multiplicity of the root $z=0$ for the polynomials $S_{k+1}$, $S_{k-1}$ is equal to $9$. Thus we have proved that every polynomial $T_k$ possesses only simple roots except possibly the point $z=0$.  As a consequence we see that the polynomial $S_k$ does not have multiple roots except possibly the point $z=0$. Indeed, assuming the converse, we see that any neighbor polynomial from the sequence $\{T_k\}$, i.e. $T_{k\pm1}$ possesses a multiple root, what is impossible. Now let us suppose that the polynomials $S_k(z)$, $T_{k\pm1}(z)$ or $S_{k\pm1}(z)$, $T_{k}(z)$ have a common root $z=z_0$, $z_0\neq0$. By the above the root $z=z_0$ is simple for all these polynomials and this fact contradicts  expressions \eqref{K2_RS}, \eqref{K2_1_RS}.

Further we see that there are polynomials in the sequences $\{S_k\}$, $\{T_k\}$ possessing the multiple root at the point $z=0$. Indeed with the help of the Laurent series around infinity that satisfy the equations for polynomials  (see \eqref{ODS_ST_1}), we find their structure
\begin{equation}
\begin{gathered}
\label{TS_1_Structre}S_k(z)=\sum_{i=0}^{[\deg S_k/5]}A_{k,i}z^{\deg S_k-5i},\quad T_k(z)=\sum_{i=0}^{[\deg T_k/5]}B_{k,i}z^{\deg T_k-5i},
\end{gathered}
\end{equation}
where the degrees of the polynomials $S_k$, $T_k$ are given in \eqref{TS_degrees}. From this expressions it follows that if $\deg S_k \mod 5\neq 0$ or $\deg T_k \mod 5\neq 0$, then the corresponding polynomial has the multiple root at the point $z=0$. Note that for a polynomial $S_k$ with multiple zero root of multiplicity $8$ the coefficient of $z^3$ in expression \eqref{TS_1_Structre} is equal to zero.

Now let us take two neighbor polynomials from the sequences $\{S_k\}$, $\{T_k\}$, for example, $S_k$, $T_{k+1}$. Suppose that the polynomial $S_k$ has a root of multiplicity $8$ at the point $z=0$ and the polynomial $T_k$ has a root of multiplicity $2$ at the point $z=0$, then we can set $S_k(z)=z^8R(z)$, $T_{k+1}(z)=z^2V(z)$, where the polynomials $R(z)$, $V(z)$ do not have multiple or common roots and $R(0)\neq0$, $V(0)\neq0$. Substituting this expressions into the relation given in \eqref{ST_Rec_rel1}, we obtain
\begin{equation}
\begin{gathered}
\label{TS_1_Additiona_relation1}z(R_{zz}V-4 R_zV_z+4RV_{zz})+8(R_zV-2 RV_z)=0
\end{gathered}
\end{equation}
This formula coincides with the expression \eqref{Correlation_for_Polynomials_mu_new} if we set $P=R$, $Q=V$, $\mu=2$, $\nu=-4$. By analogy we consider the case $S_k(z)=zR(z)$, $T_{k+1}(z)=z^2V(z)$ and obtain the relation
\begin{equation}
\begin{gathered}
\label{TS_1_Additiona_relation2}z(R_{zz}V-4 R_zV_z+4RV_{zz})-6(R_zV-2 RV_z)=0
\end{gathered}
\end{equation}
Thus we see that the formula \eqref{TS_1_Additiona_relation2} coincides with the expression \eqref{Correlation_for_Polynomials_mu_new} setting $P=R$, $Q=V$, $\mu=2$, $\nu=3$. If the point $z=0$ is not a multiple root of of the polynomial $T_{k+1}$, then the polynomials $S_k$, $T_{k+1}$ do not have multiple and common roots and satisfy the correlation, which coincides with \eqref{Correlation_for_Polynomials_mu} setting $P=S_k$, $Q=T_{k+1}$, $\mu=2$. Consequently, we have shown that roots of two neighbor polynomials from the sequences $\{S_k\}$, $\{T_k\}$, i.~e. $S_k$, $T_{k+1}$ or $S_{k+1}$, $T_{k}$  give stationary equilibrium positions of point vortices in arrangements described in section~\ref{Dynamics}.

\begin{figure}[t]
 \centerline{
 \subfigure[$S_6^{(1)}(z)$]{\epsfig{file=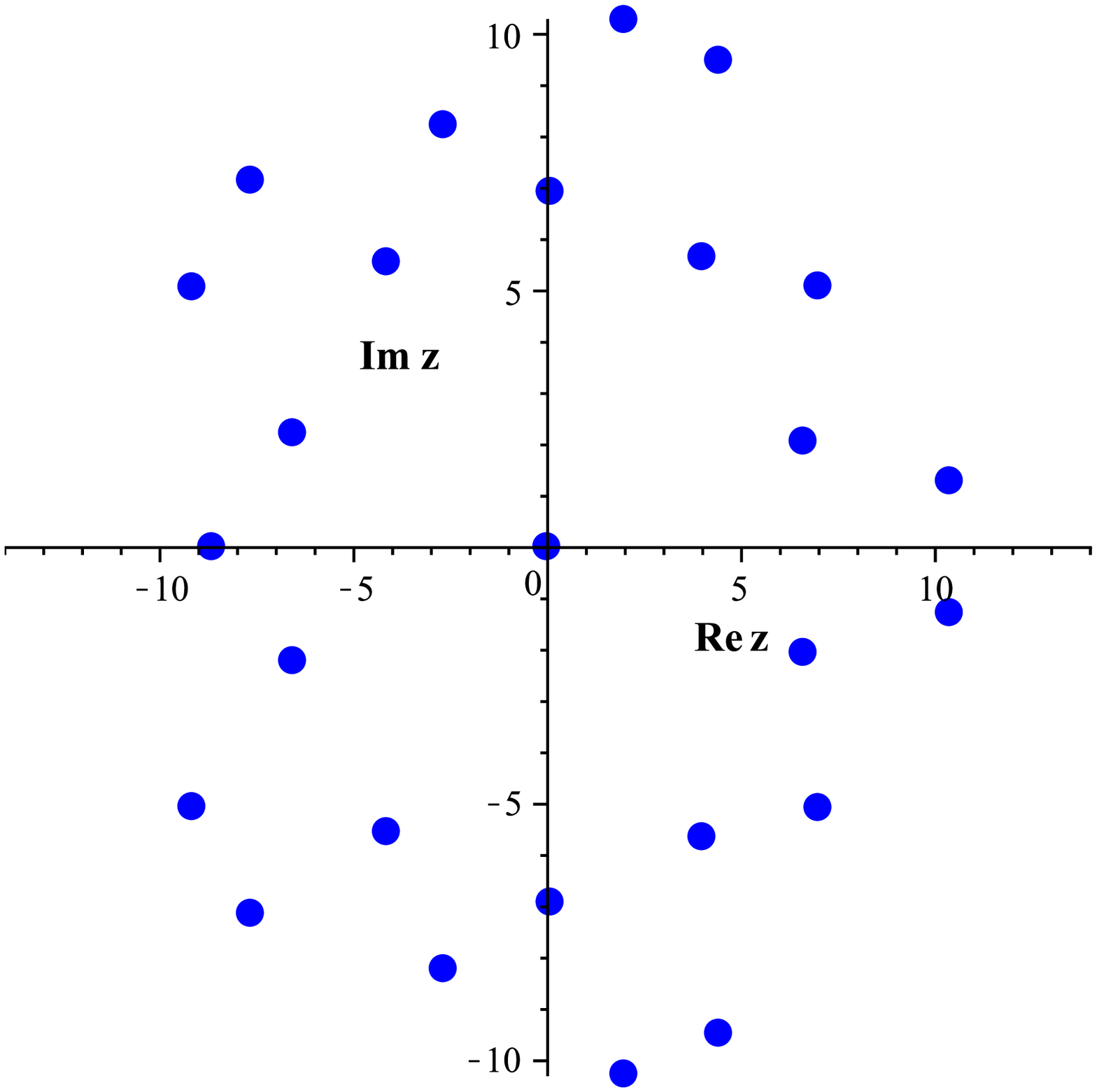,width=60mm}}
 \subfigure[$T_7^{(1)}(z)$]{\epsfig{file=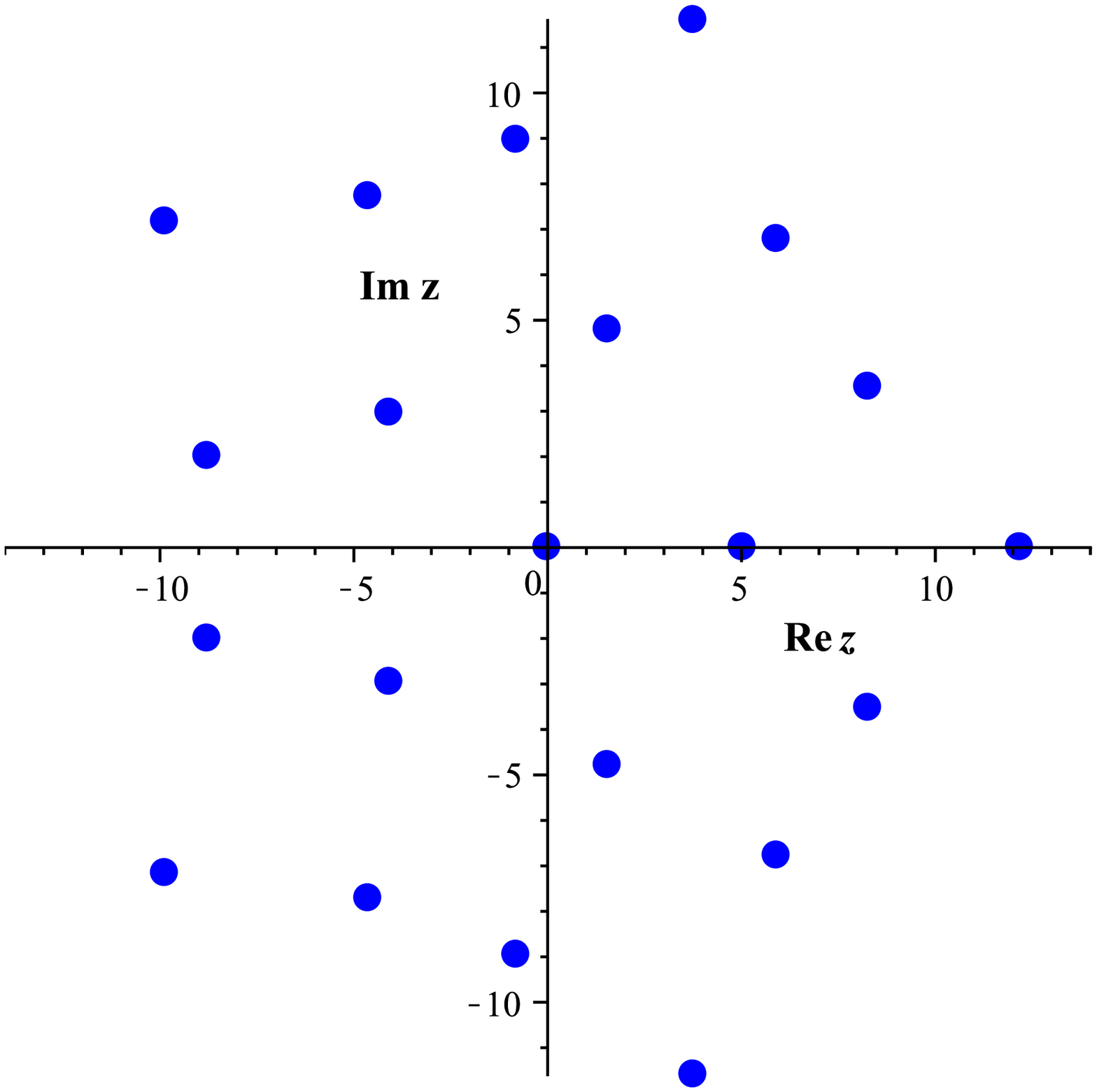,width=60mm}}}
   \caption{Roots of the polynomials $\{S_k^{(1)}(z)\}$, $\{T_k^{(1)}(z)\}$, $\alpha_1=10$.}
 \label{F:Plots_K2_1}
\end{figure}

\begin{figure}[t]
 \centerline{
   \subfigure[$S_{11}^{(1)}(z)$]{\epsfig{file=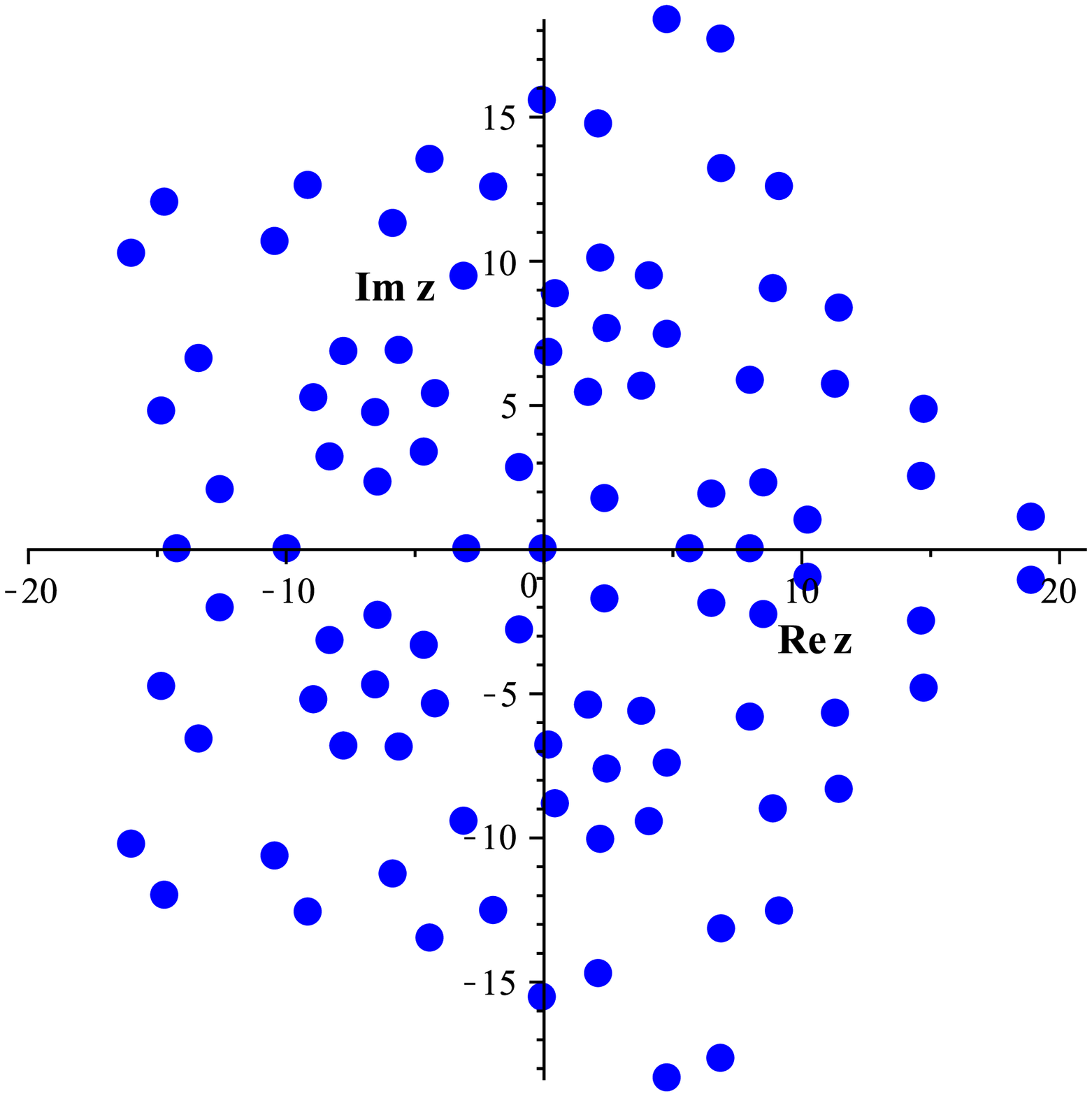,width=60mm}}
 \subfigure[$T_{12}^{(1)}(z)$]{\epsfig{file=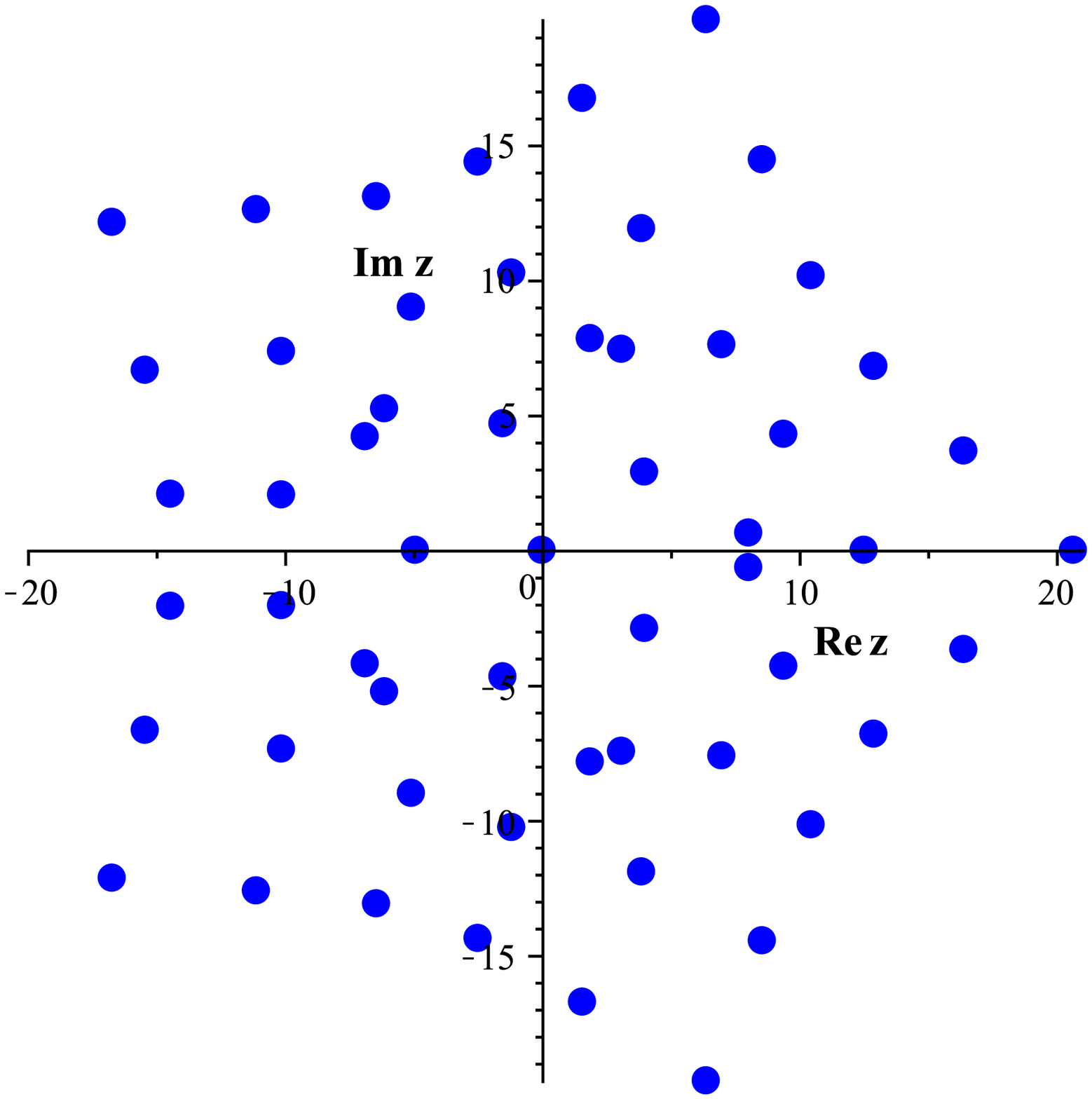,width=60mm}}}
 \caption{Roots of the polynomials $\{S_k^{(1)}(z)\}$, $\{T_k^{(1)}(z)\}$, $\alpha_1=10$.}
 \label{F:Plots_K2_2}
\end{figure}

In conclusion let us find algebraic relations satisfied by the roots of polynomials $S_k$, $T_k$. By $\{a_i\}$ we denote the roots of $S_k$ and by $\{b_j\}$ the roots of $T_k$. The functions $h=(\ln S_k)_z$, $g=(\ln T_k)_z$ can be presented in the form
\begin{equation}
\begin{gathered}
\label{TS_1_TS_poles}h(z)=\frac{\Delta_1}{z}+\sum_{i=1}^{\deg S_k-\Delta_1}\frac{1}{z-a_i},\quad g(z)=\frac{\Delta_2}{z}+\sum_{j=1}^{\deg T_k-\Delta_2}\frac{1}{z-b_j},
\end{gathered}
\end{equation}
where $\Delta_1=0$ if $S_k(0)\neq0$, $\Delta_1=1$ if the point $z=0$ is a simple root of the polynomial $S_k$, $\Delta_1=8$ if the point $z=0$ is the multiple root of the polynomial $S_k$ and in its turn $\Delta_2=0$ if $T_k(0)\neq0$, $\Delta_2=1$ if the point $z=0$ is a simple root of the polynomial $T_k$, $\Delta_2=2$ if the point $z=0$ is the multiple root of the polynomial $T_k$. We suppose that $a_i=0$, $i=\deg S_k-\Delta_1+1$, $\ldots$, $\deg S_k$ in the case $S_k(0)=0$ and $b_j=0$, $j=\deg T_k-\Delta_2+1$, $\ldots$, $\deg T_k$ in the case $T_k(0)=0$. Finding the Laurent series for the functions $h(z)$, $g(z)$ in a neighborhood of infinity, we get
\begin{equation}
\begin{gathered}
\label{TS_1_TS_poles_Laurent_infinity}h(z)=\frac{\deg S_k}{z}+\sum_{m=1}^{\infty}\left[\sum_{i=1}^{\deg S_k}a_i^m\right]\frac{1}{z^{m+1}},\quad |z|>\max_{1\leq i \leq \deg S_k}|a_i|,\hfill \\
g(z)=\frac{\deg T_k}{z}+\sum_{m=1}^{\infty}\left[\sum_{j=1}^{\deg T_k}b_j^m\right]\frac{1}{z^{m+1}},\quad |z|>\max_{1\leq j \leq \deg T_k}|b_j|. \hfill
\end{gathered}
\end{equation}
Further we take the equations satisfied  by he functions $h(z)$, $g(z)$ and construct the Laurent series in a neighborhood of infinity using asymptotic methods. Finally, we obtain
\begin{equation}
\begin{gathered}
\label{TS_1_TS_poles_Laurent_infinity_AS}h(z)=\frac{\deg S_k}{z}+\sum_{m=1}^{\infty}\frac{c_{-m-1}}{z^{m+1}},\quad g(z)=\frac{\deg T_k}{z}+\sum_{m=1}^{\infty}\frac{d_{-m-1}}{z^{m+1}}
\end{gathered}
\end{equation}
The coefficients of $z^{-1}$ in this expression are arbitrary. We take them as $\deg S_k$ for the function $h(z)$ and as $\deg T_k$ for the function $T_k$. In addition we get that $c_{-m-1}=0$ if $m \mod 5 \neq0$ and $d_{-m-1}=0$ if $m \mod 5 \neq0$. Comparing the series given by \eqref{TS_1_TS_poles_Laurent_infinity} with the series in expression \eqref{TS_1_TS_poles_Laurent_infinity_AS}, we obtain the following algebraic relations for the roots of the polynomials $S_k$, $T_k$
\begin{equation}
\begin{gathered}
\label{TS_1_TS_poles_Relations1}\sum_{i=1}^{\deg S_k}a_i^m=c_{-m-1},\quad \sum_{j=1}^{\deg T_k}b_j^m=d_{-m-1},\quad m>0.
\end{gathered}
\end{equation}
Several examples are given in table \ref{t:relations1}.
\begin{table}[h]
    \caption{Algebraic relations for the roots $\{a_i\}$ of the polynomial $S_k^{(1)}(z)$ and for the roots $\{b_i\}$ of the polynomial $T_k^{(1)}(z)$ (the upper index is omitted).} \label{t:relations1}
       \begin{tabular}[pos]{|c|c|c|c|}
    \hline
    $\displaystyle \sum_{i=1}^{\deg S_k}a_i = 0$ & $\displaystyle \sum_{i=1}^{\deg S_k}a_i^2 = 0$ & $\displaystyle \sum_{i=1}^{\deg S_k}a_i^3 = 0$ & $\displaystyle \sum_{i=1}^{\deg S_k}a_i^4 = 0$ \\
    \hline
    $\displaystyle \sum_{j=1}^{\deg T_k}b_j = 0$ & $\displaystyle \sum_{j=1}^{\deg T_k}b_j^2 = 0$ & $\displaystyle \sum_{j=1}^{\deg T_k}b_j^3 = 0$ & $\displaystyle \sum_{j=1}^{\deg T_k}b_j^4 = 0$ \\
    \hline
    \hline
    \multicolumn{4}{|c|}{$\displaystyle \sum_{i=1}^{\deg S_k}a_i^5 = 3\alpha_1\deg S_k(\deg S_k-1)(\deg S_k-8)$}\\
    \hline
    \multicolumn{4}{|c|}{$\displaystyle \sum_{j=1}^{\deg T_k}b_j^5 = 12\alpha_1\deg T_k(\deg T_k-1)(\deg T_k-2)$}\\
    \hline
        \end{tabular}
\end{table}
We note that expanding the functions $h(z)$, $g(z)$ in a neighborhood of their poles and the origin gives other sequences of algebraic relations for the roots of polynomials $S_k$, $T_k$. In figures \ref{F:Plots_K2_1}, \ref{F:Plots_K2_2} we present several plots of roots for the polynomials $S_k$, $T_k$.



\section{Conclusion}

In this article we have studied the problem of finding stationary equilibrium positions for two sets of point vortices (identical vortices with strength $\Gamma$ in one set and  identical vortices with strength $-\mu\Gamma$ in another). Along with this we have investigated the case of three sets of vortices (identical vortices with strength $\Gamma$ in the first set, identical vortices with strength $-\mu\Gamma$ in the second and a vortex in the origin with strength $-\nu \Gamma$). We have found differential equations satisfied by the generating polynomials of vortices and have derived formulae for constructing these polynomials explicitly in the case $\mu=2$. Interestingly, that if we introduce the generalized Hirota derivative according to the rule
\begin{equation}
\begin{gathered}
\label{Gen_Hirota}D_z^m[\mu]\,f(z)\cdot
g(z)=\left[\left(\frac{d}{d{z_1}}-\mu\frac{d}{d{z_2}}\right)^mf(z_1)g(z_2)
\right]_{z_1=z_2=z},\quad m\in \mathbb{N}\cup\{0\},
\end{gathered}
\end{equation}
then we can rewrite differential equations for the generating polynomials of vortices as follows
\begin{equation}
\begin{gathered}
\label{Gen_Hirota_Rel_Vor}D_z^2[\mu]P\cdot Q=0,\quad \left\{zD_z^2[\mu]-2\nu D_z[\mu]\right\}P\cdot Q=0.
\end{gathered}
\end{equation}
 Further we have shown that special polynomials associated with rational solutions of the generalized $K_2$ hierarchy satisfy the same differential equations with $\mu=2$. We have derived differential -- difference relations and ordinary differential equations satisfied by the polynomials. In details we have studied the case of the fourth--order equation in the generalized $K_2$ hierarchy. We have shown that roots of the polynomials associated with this equation give equilibrium positions of the vortices in situations described above. We have found algebraic relations satisfied by the roots of these special polynomials and, consequently, satisfied by coordinates of vortices in equilibrium.

In conclusion we would like to note that equations in the Fordy--Gibons, Sawada--Kotera, and  Kaup--Kupershmidt hierarchies possess solutions expressible via solutions of the equations in the generalized $K_2$ hierarchy. As a consequence, there exist solutions of these partial differential equations expressible in terms of special polynomials, which we have studied in this article.

\section{Acknowledgements}

We would like to thank the referee for useful comments, which helped us to improve the manuscript, and for attracting our attention to the work of Loutsenko \cite{Loutsenko01}.

This research was partially supported by Federal Target Programm
"Research and Scientific--Pedagogical Personnel of Innovation
in Russian Federation on 2009-–2013".

\end{document}